\documentclass[a4paper]{PoS}

\title{X-ray Polarimetry - a Tool for the Galactic center diagnosis}

\ShortTitle{X-ray Polarimetry \& the Galactic center }

\author{\speaker{Fr\'ed\'eric~Marin}\\
        Astronomical Institute of the Academy of Sciences, Bo{\v c}n\'{\i} II 1401, CZ-14100 Prague, Czech Republic\\
        E-mail: \email{frederic.marin@asu.cas.cz}}

\abstract{Was the Milky Way galaxy a low-luminosity active galactic nucleus (AGN) in the past?
	  Can we find traces of remnant structures supporting this idea? What is the 
	  three-dimensional arrangement of matter around our central supermassive black 
	  hole? A number of fundamental questions concerning our own Galactic center
	  remain controversial. To reveal the structure of the high-energy sky around our 
	  galactic core, a technique more sensitive to the morphology of the emitters than 
	  spectroscopy is needed. In this lecture note, I describe how X-ray polarimetry 
	  can open a new observational window by precisely measuring the three-dimensional 
	  position of the scattering material in the Galactic center. The observed polarization 
	  degree and polarization position angle would also determine unambiguously the primary 
	  source of emission and trace the centennial history of our supermassive black hole by 
	  detecting echoes of its past activity thanks to astrophysical mirrors. Finally, the 
	  synergy between X-ray polarimetry and infrared and radio observations can be used to 
	  better constrain the geometry in the first hundred parsecs around the Galactic center.}

\FullConference{XI Multifrequency Behaviour of High Energy Cosmic Sources Workshop\\
		25-30 May 2015\\
		Palermo, Italy}

\begin{document}

\section{The Galactic center region}
At a distance of 8.33~$\pm$~0.35  kpc \cite{Gillessen2009} from our Solar system is an astrophysical
environment that presents an extremely rich collection of gaseous, stellar and gravitational components: 
the Galactic center (GC). Due to the GC proximity, the angular size of its central supermassive black 
hole (9.65~$\mu$as) may become resolvable in the near future thanks to very-long-baseline interferometry 
in the radio domain, making the Milky Way an excellent laboratory to locally investigate the processes 
happening in more distant galaxies. Until then, observations are studying Galactic regions up to a 
few hundred parsecs from the GC. 

Within this Galactic volume, the stellar population is rich in young and massive stars, which is 
characteristic of a recent and/or ongoing star formation. The stars are essentially concentrated in
a disk-like structure called the nuclear stellar disk, extending up to 230~pc in width and 45~pc in 
height; with stellar clusters being more frequent in the center of the nuclear bulge \cite{Launhardt2002}. 
At similar distance scales, the maximum concentration of interstellar gas is gathered in an inclined 
molecular and atomic disk that extends up to 1.5 kpc from the center, carrying about 10$^7$~M$_\odot$ of H~I
\cite{Burton1978}. This disk reaches the inner 2~pc of the GC, where it forms a non axisymmetric reservoir 
of gas with total mass of 2 -- 5 $\times$ 10$^5$~M$_\odot$ \cite{Oka2011}. This dense, warm, molecular and atomic 
region is seen nearly edge-on ($\le$ 20$^\circ$), extends up to 3~pc along the North-East direction and up 
to 7~pc along the South-West \cite{Genzel1985}, and is usually denoted as the circumnuclear disk (CND). 
More generally, the first hundred parsecs around the CND is rich in molecular gas, so that this region 
is named the central molecular zone (CMZ). The central two parsecs of the GC are occupied by the Sgr~A West
region which appears to be dominated by ionized gas emission. Finally, at the very center is a radio-intense, 
sub-arcsecond, structure that has been discovered in 1974 \cite{Balick1974} and later associated with 
a strong gravitational field that governs the orbit of nearby stars \cite{Eckart2002}. This is considered 
as the first reliable identification of a supermassive black hole.

The GC is thus a vast concentration of molecular and atomic gaseous regions, bright star clusters, 
dynamical structures and, more importantly, a central supermassive black hole (Sgr~A$^*$). Its proximity allows
us to probe the direct environment of a $>$~10$^5$~M$_\odot$ compact region and better constrain the growth 
mechanism of black holes, the physical condition of the early Universe and the formation of galaxies.

\section{X-ray polarimetry as a new observational window}

\begin{figure*}
   \centering
   \includegraphics[trim = 0mm 25mm 0mm 30mm, clip, width=14cm]{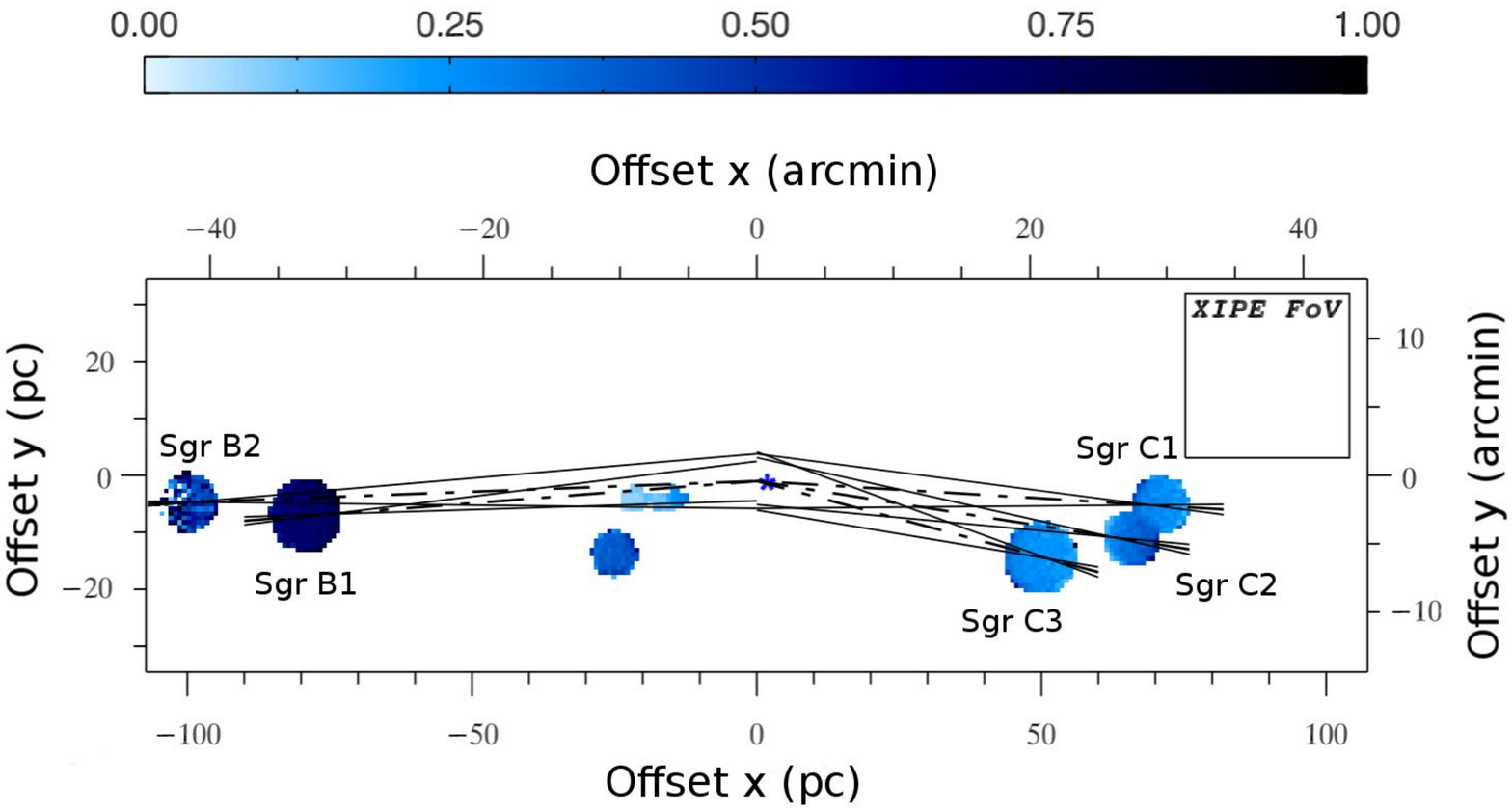}
   \caption{Simulated 4 -- 8~keV polarimetric map of the GC, using the point spread function 
	    of the {\it XIPE} mission. The polarization degree is color-coded and ranges 
	    from white (0~\%) to blue (100~\%). Eleven of the most observed reflection 
	    nebulae has been simulated to observe the effect of scattering-induced linear 
	    polarization. Dark segments are representative of the estimated polarization 
	    position angle (dashed line) and its associated error (solid line).
	    The field of view (FoV) of {\it XIPE} is indicated with a white box, and a 
	    blue star indicates the position of Sgr~A$^*$.}
   \label{Fig:XIPE}
\end{figure*}

Despite its proximity and the numerous radio to $\gamma$-ray observations of the GC, the history of 
the central parsecs of the Milky Way remains questioned. The quiescence of Sgr~A$^*$, which accretes at a 
fairly slow rate (10$^{-8}$ M$_{\odot}$.y$^{-1}$ near the event horizon \cite{Baganoff2003}), is
in contradiction with past observations of nearby nebulae \cite{Sunyaev1993,Murakami2000}. 
Situated at projected distances $<$~150~pc, a collection of molecular gas clouds has shown 
to have ``strong'' ($L_X~\sim$~10$^{35}$~ergs s$^{-1}$ \cite{Murakami2000}) absorption-corrected 
X-ray luminosities in comparison with Sgr~A$^*$ ($\sim$~2.4~$\times$~10$^{33}$~ergs s$^{-1}$ 
\cite{Baganoff2003}). Their spectra can be described by a power-law continuum and a prominent 6.4~keV 
iron emission line, both being attributed to Compton scattering and fluorescence processes. The 
hypothesis of an irradiating X-ray source located inside the nebulae has been ruled out 
by the absence of significant long-term variability of the iron K$\alpha$ line \cite{Revnivtsev2004}. 
In addition, the observed disagreement between the molecular mass distribution of the Sgr~B2 cloud 
and the location of the peak of the extended 6.4 keV emission suggests an external source that 
would be located in the vicinity of Sgr~A$^*$ \cite{Murakami2000}. One possible conclusion is 
that the 6.4~keV bright, giant molecular clouds are reprocessing a past X-ray activity of Sgr~A$^*$, 
of which the duration and intensity are debated.

To evaluate the intensity of the activity period (not to be mistaken with short-lived X-ray flares
nowadays observed in the GC, e.g. \cite{Porquet2003}), one can relate the intrinsic luminosity of 
the source to the iron K$\alpha$ line flux and to the inverse of the distance to Sgr~A$^*$. If the 
former is known (with typical uncertainties lower than 15~\%), the latter is restricted to the 
observed projected distance between the supermassive black hole and the reflection nebula. 
X-ray spectroscopy \cite{Ryu2009} and radio observations \cite{Sawada2004} were used to infer 
the position of the scattering nebulae but their results suffer from large uncertainties that does
not allow to precisely constraint the solid angle of the cloud from the location of the primary 
source. This is a crucial piece of information (together with the cloud optical depth)
to properly constrain the history and morphology of the GC.

Information about the spatial location of the last scattering event are encoded in photons
by a phenomenon called polarization. In white light, most sources of light are incoherent and 
randomly polarized (over time the polarization is constantly changing in an unpredictable manner), 
but scattering can force the electric field vector to oscillate in a given plane, resulting in 
measurable polarization. Reflection nebulae are thus potential targets for X-ray polarimetry. 
Moreover, the degree of scatter-induced polarization is strongly correlated with the angle 
between the source, the mirror and the observer. Based on these properties, it has been proved 
that the molecular cloud Sgr~B2 should produce large ($>$~30~\%) degrees of linear polarization 
in the 2 -- 8~keV \cite{Churazov2002} and 8 -- 35~keV \cite{Marin2014a,Marin2014b} regimes, 
if the primary source of emission is Sgr~A$^*$. Using more advanced simulations, a 4 -- 8~keV 
polarization map of the 100~$\times$~30~pc around Sgr~A$^*$ has been produced (see Fig.~\ref{Fig:XIPE} 
and \cite{Marin2015}). The resulting polarization of the eleven most observed/constrained reflection 
nebulae (Sgr~B1, Sgr~B2, G0.11-0.11, Bridge~E, Bridge~D, Bridge~B2, MC2, MC1, Sgr~C3, Sgr~C2, and 
Sgr~C1) presents a variety of signatures, ranging from nearly unpolarized (The Bridge) to highly 
polarized ($\sim$ 77\% for the Sgr~B complex) fluxes, with their polarization position angle normal 
to the scattering plane. Despite strong dilution due to a diffuse plasma emission\footnote{This 
two-temperature plasma is probably due to a multitude of faint sources (accreting white dwarfs and 
coronally active stars \cite{Revnivtsev2007}).} angularly superimposed to the X-ray emission of the 
reflection nebulae, the Sgr~B and Sgr~C complexes, and the G0.11-0.11 cloud, have proven to be in 
the detectability range of a modern X-ray polarimeter. The X-ray Imaging Polarimetry Explorer 
({\it XIPE} \cite{Soffitta2013}), a mission concept selected as a candidate for the ESA's next 
medium-class science mission, is a prototypical satellite that could detect the polarization of these 
reflection nebulae. With a 3~Ms long observation of the Sgr~B and C complexes, both the degree of 
polarization (related to the three-dimensional position of the cloud) and the polarization position 
angles (pinpointing the angular position of the illuminating source) would be detectable \cite{Marin2015}.

\section{A multi-wavelength complementarity}

\begin{figure*}
   \centering
   \includegraphics[trim = 0mm 10mm 0mm 15mm, clip, width=12.5cm]{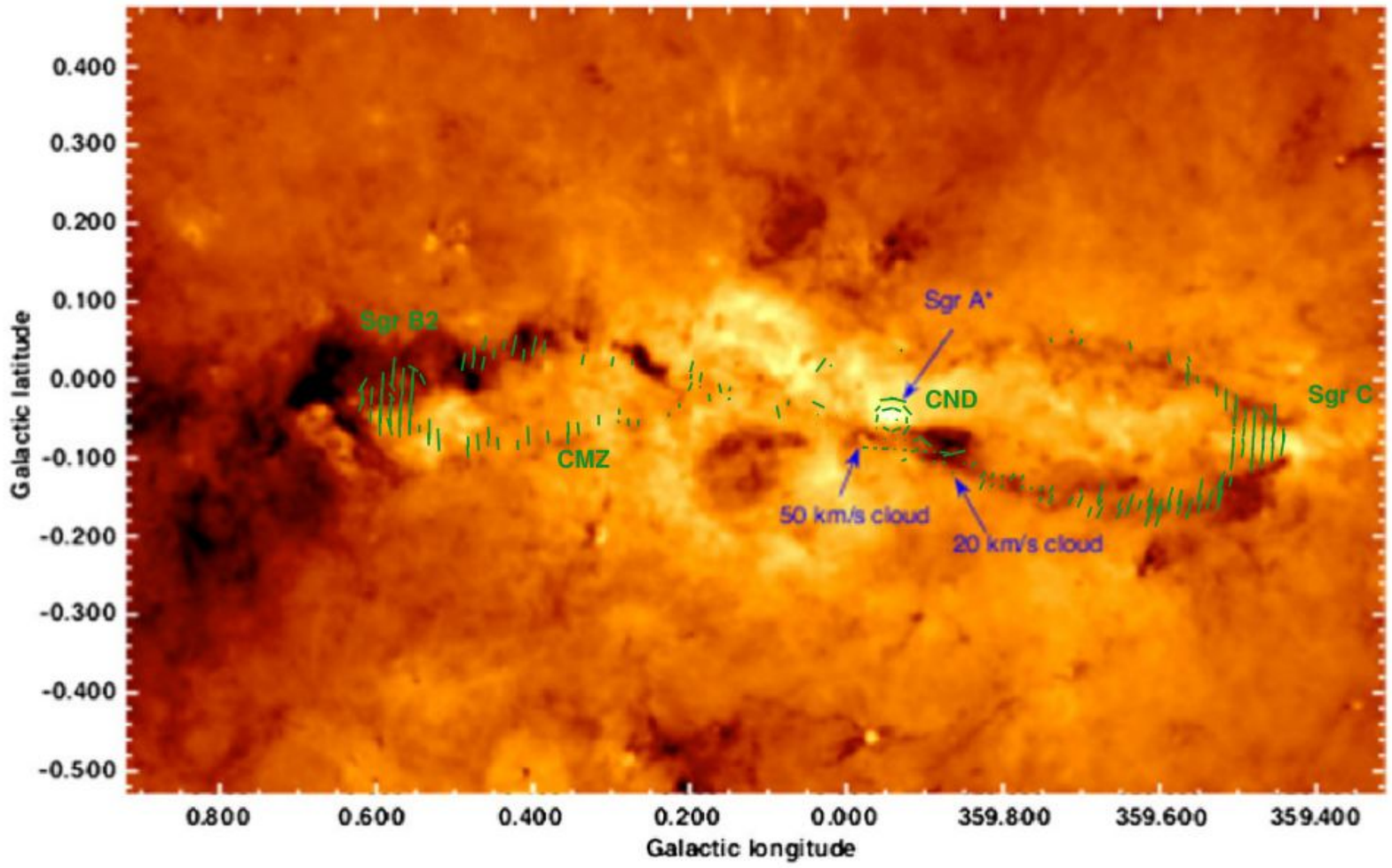}
   \caption{An example of broadband synergy between IR spectroscopy (temperature map from \cite{Molinari2011}) 
	    and X-ray polarimetry (green lines, representative of the polarization degree and polarization position 
	    angle \cite{Marin2014a}) in the Galactic center. The central, quiescent, SMBH Sgr~A* is revealed by 
	    its imprints onto the central molecular zone (CMZ), its circumnuclear disc (CND) and on the reflection 
	    nebulae (Sgr~B and C complexes).}
   \label{Fig:complementarity}
\end{figure*}

Thanks to X-ray polarimetry, the spatial location of the reflection nebulae with respect to the 
primary source can be deduced. Additionally, measuring the polarization position angle would 
unambiguously point towards the primary source (see Fig.~\ref{Fig:XIPE}), giving proof or rejecting 
the hypothesis that the GC was active in the past (with an activity similar to a low-luminosity AGN).

Traces of this past activity are difficult to assess at other wavebands. One of the most promising
hints is a continuous chain of irregular clumps around the Galactic core that has been revealed by the 
far-infrared cameras on-board of {\it Herschel} \cite{Molinari2011}. This elliptical twisted ring of molecular 
material is part of the CMZ, and extends up to $\sim$~100~pc in longitude and $\sim$~25~pc in latitude. 
The coherence of the ring's morphology indicates stability over time and it is reminiscent of the 
dusty tori surrounding the central regions of AGN. A LMT/AzTEC 1.1 mm survey of the CMZ (D.~Wang, 
private communication) confirms the existence of this peculiar structure. Completely optically 
thin to dust emission, the conjugation of this radio map with maps at shorter wavelengths will allow 
to put strong dust-based constraints on the mass of dense clouds. Together with multi-band near-infrared 
images, it might be possible to determine the line-of-sight locations of the clouds based on extinction 
measurements. Until then, the geometrical size of the CMZ, its column density in excess of 
10$^{24}$~cm$^{-2}$, and its orbital speed ($\sim$100 km s$^{-1}$), are all compatible with AGN tori 
\cite{Shi2006}. 

As an illustration of the power of X-ray polarimetry in combination with other wavelengths, results 
from 8 -- 35~keV polarimetric simulations of the GC \cite{Marin2014a}, including the Sgr~B and C complexes, 
the CND, and the $\infty$-shaped CMZ, have been superimposed on the temperature map from \cite{Molinari2011} 
in Fig.~\ref{Fig:complementarity}. The CMZ polarization follows the morphology of the twisted structure, 
with partial disappearance of the polarization vectors due to low polarization degrees induced by 
unfavorable scattering geometries and dilution from the unpolarized primary flux. Maximum polarization 
is expected at the Eastern and Western quadratures, where the CMZ mixes with the giant molecular clouds. 

{\small \acknowledgments
I would like to acknowledge my collaborators Fabio Muleri, Paollo Soffitta, Vladim\'ir Karas and Devaky Kunneriath 
for the work accomplished together in the topic of X-ray polarimetry. I am also grateful to Sergio Molinari who 
kindly accepted to share his 2011's temperature map of the Galactic center region, Michal {\v S}vanda who helped
me to create the IR/X-ray map, and to Daniel Wang for his insights on the radio domain.}


\begin{thebibliography}{99}
\bibitem{Baganoff2003} Baganoff, F.~K., Maeda, Y., Morris, M., et al.,
\emph{Chandra X-Ray Spectroscopic Imaging of Sagittarius A* and the Central Parsec of the Galaxy},
\emph{ApJ} {\bf 591} (891)
[{\tt arXiv:astro-ph/0102151}].

\bibitem{Balick1974} Balick, B., \& Brown, R.~L.,
\emph{Intense sub-arcsecond structure in the galactic center},
\emph{ApJ} {\bf 194} (265).

\bibitem{Bender2005} Bender, R., Kormendy, J., Bower, G., et al.,
\emph{HST STIS Spectroscopy of the Triple Nucleus of M31: Two Nested Disks in Keplerian Rotation around a Supermassive Black Hole},
\emph{ApJ} {\bf 631} (280) 
[{\tt arXiv:astro-ph/0509839}].

\bibitem{Bettoni2003} Bettoni, D., Falomo, R., Fasano, G., \& Govoni, F.,
\emph{The black hole mass of low redshift radiogalaxies},
\emph{A\&A} {\bf 399} (869) 
[{\tt arXiv:astro-ph/0212162}].

\bibitem{Burton1978} Burton, W.~B., \& Liszt, H.~S.,
\emph{The gas distribution in the central region of the Galaxy. I - Atomic hydrogen},
\emph{ApJ} {\bf 225} (815).

\bibitem{Churazov2002} Churazov, E., Sunyaev, R., \& Sazonov, S.,
\emph{Polarization of X-ray emission from the Sgr B2 cloud},
\emph{MNRAS} {\bf 330} (817)
[{\tt arXiv:astro-ph/0111065}].

\bibitem{Eckart2002} Eckart, A., Genzel, R., Ott, T., \& Sch{\"o}del, R.,
\emph{Stellar orbits near Sagittarius~A$^*$},
\emph{MNRAS} {\bf 331} (917)
[{\tt arXiv:astro-ph/0201031}].

\bibitem{Genzel1985} Genzel, R., Crawford, M.~K., Townes, C.~H., \& Watson, D.~M.,
\emph{The neutral-gas disk around the galactic center},
\emph{ApJ} {\bf 297} (766).

\bibitem{Gillessen2009} Gillessen, S., Eisenhauer, F., Trippe, S., et al.,
\emph{Monitoring Stellar Orbits Around the Massive Black Hole in the Galactic Center},
\emph{ApJ} {\bf 692} (1075) 
[{\tt arXiv:0810.4674}].

\bibitem{Launhardt2002} Launhardt, R., Zylka, R., \& Mezger, P.~G.,
\emph{The nuclear bulge of the Galaxy. III. Large-scale physical characteristics of stars and interstellar matter},
\emph{A\&A} {\bf 384} (112) 
[{\tt arXiv:astro-ph/0201294}].

\bibitem{Marin2014a} Marin, F., Karas, V., Kunneriath, D., \& Muleri, F.,
\emph{Prospects of 3D mapping of the Galactic Centre clouds with X-ray polarimetry},
\emph{MNRAS} {\bf 441} (3170)
[{\tt arXiv:1405.0898}].

\bibitem{Marin2014b} Marin, F., Karas, V., Kunneriath, D., Muleri, F., \& Soffitta, P.,
\emph{Probing the Galactic center with X-ray polarimetry},
\emph{Proc. of the SF2A} {\bf 109}
[{\tt arXiv:1408.0354}].

\bibitem{Marin2015} Marin, F., Muleri, F., Soffitta, P., Karas, V., \& Kunneriath, D.,
\emph{Reflection nebulae in the Galactic center: soft X-ray imaging polarimetry},
\emph{A\&A} {\bf 576} (A19)
[{\tt arXiv:1502.04894}].

\bibitem{Molinari2011} Molinari, S., Bally, J., Noriega-Crespo, A., et al., 
\emph{A 100 pc Elliptical and Twisted Ring of Cold and Dense Molecular Clouds Revealed by Herschel Around the Galactic Center},
\emph{ApJL} {\bf 735} (L33) 
[{\tt arXiv:1105.5486}].

\bibitem{Murakami2000} Murakami, H., Koyama, K., Sakano, M., Tsujimoto, M., \& Maeda, Y.,
\emph{ASCA Observations of the Sagittarius B2 Cloud: An X-Ray Reflection Nebula},
\emph{ApJ} {\bf 534} (283)
[{\tt arXiv:astro-ph/9908229}].

\bibitem{Oka2011} Oka, T., Nagai, M., Kamegai, K., \& Tanaka, K.,
\emph{A New Look at the Galactic Circumnuclear Disk},
\emph{ApJ} {\bf 732} (120).

\bibitem{Porquet2003} Porquet, D., Predehl, P., Aschenbach, B., et al.,
\emph{XMM-Newton observation of the brightest X-ray flare detected so far from Sgr~A$^*$},
\emph{A\&A} {\bf 407} (L17)
[{\tt arXiv:astro-ph/0307110}].

\bibitem{Revnivtsev2004} Revnivtsev, M.~G., Churazov, E.~M., Sazonov, S.~Y., et al.,
\emph{Hard X-ray view of the past activity of Sgr A* in a natural Compton mirror},
\emph{A\&A} {\bf 425} (L49)
[{\tt arXiv:astro-ph/0408190}].

\bibitem{Revnivtsev2007} Revnivtsev, M., Vikhlinin, A., \& Sazonov, S.,
\emph{Resolving the Galactic X-ray background},
\emph{A\&A} {\bf 473} (857)
[{\tt arXiv:astro-ph/0611952}].

\bibitem{Ryu2009} Ryu, S.~G., Koyama, K., Nobukawa, M., Fukuoka, R., \& Tsuru, T.~G.,
\emph{An X-Ray Face-on View of the Sgr B Molecular Clouds Observed with Suzaku},
\emph{PASJ} {\bf 61} (751)
[{\tt arXiv:0904.4550}].

\bibitem{Sawada2004} Sawada, T., Hasegawa, T., Handa, T., \& Cohen, R.~J.,
\emph{A Molecular Face-on View of the Galactic Centre Region},
\emph{MNRAS} {\bf 349} (1167)
[{\tt arXiv:astro-ph/0401286}].

\bibitem{Shi2006} Shi, Y., Rieke, G.~H., Hines, D.~C., et al.,
\emph{9.7 $\mu$m Silicate Features in Active Galactic Nuclei: New Insights into Unification Models},
\emph{ApJ} {\bf 653} (127)
[{\tt arXiv:astro-ph/0608645}].

\bibitem{Soffitta2013} Soffitta, P., Barcons, X., Bellazzini, R., et al.,
\emph{XIPE: the X-ray imaging polarimetry explorer},
\emph{Experimental Astronomy} {\bf 36} (523)
[{\tt arXiv:1309.6995}].

\bibitem{Sunyaev1993} Sunyaev, R.~A., Markevitch, M., \& Pavlinsky, M.,
\emph{The center of the Galaxy in the recent past - A view from GRANAT},
\emph{ApJ} {\bf 407} (606).
\end{thebibliography}
\end{document}